\documentclass[12pt]{iopart}
\usepackage{color}
\usepackage{iopams}  
\usepackage{graphicx}

\begin{document}

\title[Bulk viscosity in the nonlinear and anharmonic regime]
{Bulk viscosity in the nonlinear and anharmonic regime of strange quark matter}

\author{Igor A. Shovkovy$^1$ and Xinyang Wang$^2$}

\address{$^1$ Department of Applied Sciences and Mathematics, Arizona State University, Mesa, Arizona 85212, USA}

\address{$^2$ Department of Physics, Arizona State University, Tempe, Arizona 85287, USA}
\ead{igor.shovkovy@asu.edu {\rm and} xwang176@asu.edu}

\begin{abstract}
The bulk viscosity of cold, dense three-flavor quark matter is studied as a function of temperature 
and the amplitude of density oscillations. The study is also extended to the case of two different 
types of anharmonic oscillations of density. We point several qualitative effects due to the 
anharmonicity, although quantitatively they appear to be relatively small. We also find that, 
in most regions of the parameter space, with the exception of the case of a very large amplitude 
of density oscillations (i.e. $10\%$ and above), nonlinear effects and anharmonicity have a
small effect on the interplay of the nonleptonic and semileptonic processes in the bulk viscosity.
\end{abstract}



\section{Introduction}

Cold, dense quark matter is one of possible states of baryonic matter formed at very high densities.
While there is little doubt that such quark matter can be formed in principle, the value of the critical density 
needed remains unknown. This is one of the reasons why it is not settled whether such form of matter 
can exist inside neutron stars. Indeed, the interior region of neutron stars is the most likely place to 
find dense quark matter. The corresponding densities reach up to about 10 times the nuclear 
saturation density, while the temperatures remain moderately low, of the order of $1~\mbox{MeV}$ 
or less.

A recent report on a precision measurement of the mass of the binary millisecond pulsar J1614-2230 
\cite{Demorest:2010bx} seems to strongly constraint the possibility of stellar dense quark matter \cite{Ozel:2010bz}. 
Nevertheless, it is hard to rule out quark matter solely based on the measurement of such a global property 
of a star such as its mass.. More informative probes of the state of matter at the highest densities are likely 
to be based on stellar characteristics determined by the transport properties of matter, various emission rates, 
as well as certain thermodynamic properties, which are sensitive to the spectrum of quasiparticles 
at the Fermi surface and which, therefore, can provide a deeper insight into the microscopic nature 
of dense matter. 

In this paper, we study one of such transport characteristics of dense quark matter, the bulk viscosity. 
In general, viscosity is one of possible mechanisms responsible for damping of the so-called r-mode 
instabilities in compact stars \cite{Madsen:1992sx,Lindblom:1998wf,Owen:1998xg,Lindblom:1999yk,Madsen:1999ci}. 
The emission of gravitational waves tends to drive a differential collective motion of the stellar fluid in 
the form of r-modes \cite{Chandrasekhar:1992pr,Friedman:1978hf,Andersson:1997xt,Friedman:1997uh}. 
Such modes in turn increase the stellar quadrupole moment and feed back to result in even stronger 
gravitational emission. If not damped by dissipative processes, the resonant growth of the r-modes 
can eventually lead to a breakup of the star. 

\begin{figure}[b]
\noindent
\hbox{\includegraphics[width= 0.28\textwidth]{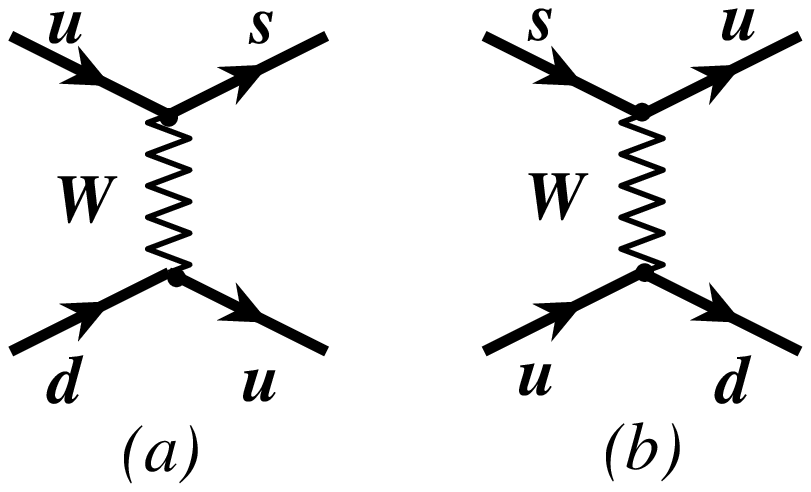}\qquad
\includegraphics[width= 0.28\textwidth]{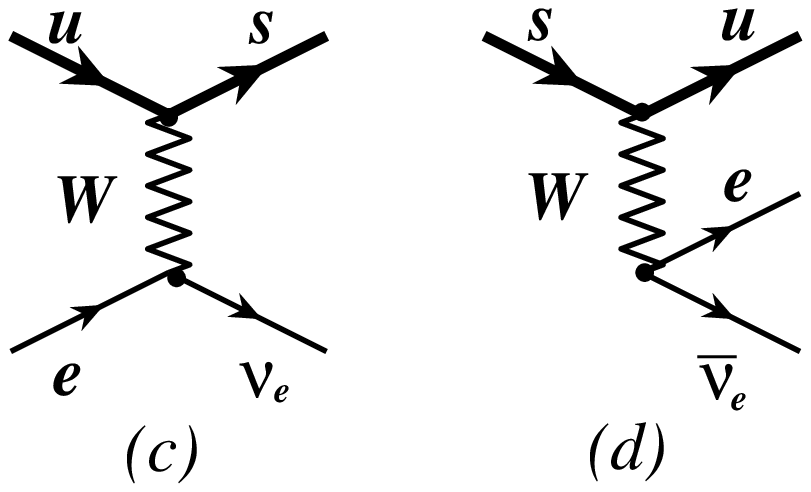}\qquad
\includegraphics[width= 0.28\textwidth]{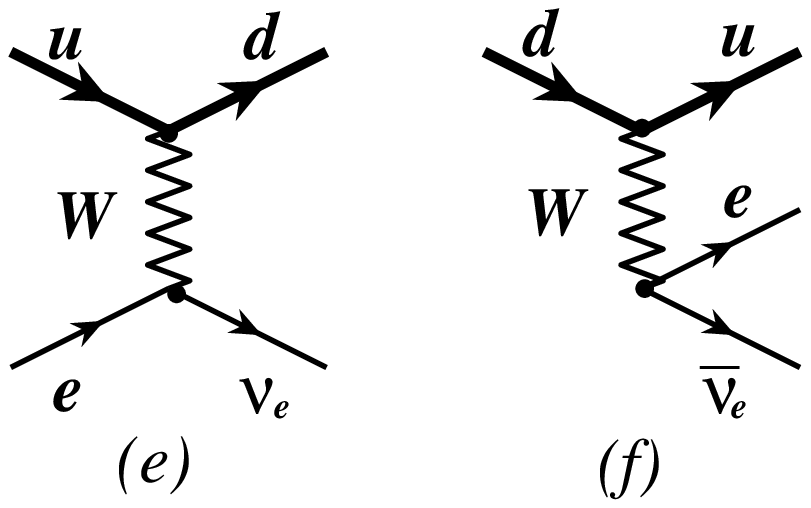}}
\caption{Diagrammatic representation of the weak processes
that contribute to the bulk viscosity of dense quark matter.}
\label{fig-weak}
\end{figure}

The bulk viscosity in the normal phase of three-flavor quark matter is usually dominated by 
nonleptonic weak processes \cite{Madsen:1992sx,Madsen:1999ci,WangLu,Sawyer,Zheng,
Xiaoping,Dong:2007mb,Huang:2009ue,Alford:2010gw}, shown in Figs.~\ref{fig-weak} (a) and (b). 
Under certain conditions, semileptonic processes, see Figs.~\ref{fig-weak} (c)--(f), can lead to an 
order of magnitude increase of the viscosity as a result of their subtle interplay with the nonleptonic 
processes \cite{Sa'd:2007ud,Wang:2010ydb}.

When the temperature is sufficiently low or when the magnitude of the density oscillation is sufficiently large, 
the rates of the weak processes may get substantial nonlinear dependence on the parameter $\delta\mu_i /T$,
where $\delta\mu_i $ are the chemical potential imbalances that measure the departure of quark matter from $\beta$ 
equilibrium. In the context of the nonleptonic processes in strange quark matter, such effects have already 
been studied \cite{Madsen:1992sx, Xiaoping,Alford:2010gw} (see also Ref.~\cite{Reisenegger:2003pd}). 
Here we extend the analysis to include also the semileptonic processes with the nonlinear corrections in 
the rates. 

When the nonlinear regime is realized, it may be reasonable to expect additional complications in the 
dynamics responsible for the energy dissipation of the hydrodynamic flow. In this paper we address one of such 
complications that is associated with the anharmonicity of the density oscillations driven by the r-modes. 
To the best of our knowledge, previously this has not been addressed in the literature. In studies of the 
bulk viscosity of stellar matter, one commonly assumes that the oscillations are perfectly harmonic. 
Of course, this is justified in the linear regime, when different harmonics do not interfere. 

In general, large magnitude density oscillations can hardly remain perfectly harmonic because higher 
harmonics can be generated by the nonlinearity. As discussed in Refs.~\cite{nonlin1996,nonlin2001a,
nonlin2001b,Schenk:2001zm,Arras:2002dw,nonlin2009}, the nonlinear regime in compact stars may be 
responsible for several qualitatively new features in the dynamics, e.g., coupling of eigenmodes, a 
non-sinusoidal shape of oscillations, various resonance phenomena, and the formation of shocks 
and turbulence. At the fundamental level, the nonlinear effects are the consequence of the 
equation of state of matter and the Einstein equations, e.g., see Ref.~\cite{nonlin2009}. Usually, they 
play a profound role when the interaction energy is comparable to the energies of modes. However, 
the nonlinearity is present in principle even at arbitrarily small amplitudes of oscillations.
Then, one of the natural questions about the nonlinear regime is: How does the associated 
non-sinusoidal shape of oscillations (i.e. the anharmonicity due to mode coupling) change the 
bulk viscosity? It is the goal of this paper to make a first attempt in exploring this issue. Our study 
will be limited to the case when the density fluctuations are much smaller than the equilibrium 
density, although they can be much larger than the temperature. In view of this limitation, we
cannot address many features of dynamics that may develop in the extreme nonlinear 
regime. Also, instead of considering realistic nonlinear oscillations found in stellar simulations \cite{nonlin1996,
nonlin2001a,nonlin2001b,Schenk:2001zm,Arras:2002dw,nonlin2009}, we will simply model the 
shape of density oscillations by using classical solutions to an anharmonic oscillator problem. 
Such a toy-model study will hopefully lead to a better understanding of the problem and will 
trigger further developments in the future.

The rest of the paper is organized as follows. In the next section, we present the general formalism for 
calculating the bulk viscosity in dense quark matter with multiple active weak processes and anharmonicity 
included. In Sec.~\ref{harmonic-regime}, we start our study of the bulk viscosity in the harmonic regime of density 
oscillations. Our results qualitatively reproduce the earlier results of Refs.~\cite{Madsen:1992sx,Alford:2010gw}
and extend them to include additional effects due to semileptonic weak processes. In the same section,
we also comment on the linear regime when the expression for the bulk viscosity can be obtained 
analytically. The anharmonic regime is studied in Sec.~\ref{anharmonic-regime} for two different 
types of density oscillations, modeled by solutions to a classical anharmonic oscillator with cubic
and quartic potentials, respectively. A brief discussion of our result is given in Sec.~\ref{discussion}.
General periodic solutions in the case of classical anharmonic oscillators with two different types of
potentials are presented in analytical form in \ref{AnharmonicOscillator}.

\section{Formalism}
\label{Formalism}

One can calculate the bulk viscosity $\zeta$ under conditions realized in stars by comparing 
the hydrodynamic relation for the energy dissipation, averaged over one period $\tau$, 
\begin {equation}
\langle \dot{\cal E}_{\rm diss}\rangle =-\frac{\zeta}{\tau} 
\int_0^{\tau} \left(\nabla \cdot \vec v\right)^2 dt 
\simeq -\frac{\zeta}{n_0^2 \tau} \int_0^{\tau} \left( \delta \dot{n} \right)^2 dt .
\label{epsilon-kin}
\end{equation}
with the thermodynamic relation for the mechanical work, counteracting the hydrodynamic flow,
\begin{equation}
\langle \dot{\cal E}_{\rm diss}\rangle 
= \frac{n}{\tau} \int_0^{\tau} P \dot{V} dt 
\simeq -\frac{1}{n_0 \tau} \int_0^{\tau} P \delta \dot{n} \, dt .
\label{diss-energy}
\end{equation}
The latter is given in terms of the instantaneous pressure $P$ and the specific volume
$V\equiv 1/n$. In the above expressions, $n_0$ is the equilibrium density and $\delta n= n-n_0$ 
is the density deviation from the equilibrium value. (Even when the magnitude of 
density oscillations is not vanishingly small, we will assume that $|\delta n| \ll n_0$.)

If the magnitude of the density oscillations $\delta n_{0}$ is vanishingly small, one
may simulate the collective motion as a harmonic oscillation, $\delta n(t) = \delta n_{0} 
\cos(\omega t)$. However, in a resonance regime, when large density oscillations develop, 
nonlinear effects may start to play an important role. In this paper, we study this 
possibility by simulating two different types of anharmonic density oscillations. 
The two types correspond to oscillators with cubic and quartic terms in the potential 
energy, which have different symmetry properties under $\delta n\to -\delta n$. The 
corresponding equations of motion read:
\begin {eqnarray}
\delta \ddot{n} +\omega_0^2 \delta n \left(1+\alpha \delta n\right)=0, &\qquad& \mbox{(Type I)}, 
\label{TypeI}\\
\delta \ddot{n} +\omega_0^2 \delta n \left(1+\beta \delta n^2\right)=0, &\qquad&  \mbox{(Type II)}.
\label{TypeII}
\end{eqnarray}
Note that the coupling constants $\alpha$ and $\beta$ have the dimensions of an inverse density and 
an inverse density squared, respectively. It is convenient, therefore, to introduce the dimensionless 
parameters $\alpha^{*}\equiv \alpha\delta n_0$ and $\beta^{*}\equiv \beta(\delta n_0)^2$, which 
are given in terms of the amplitude of density oscillations $\delta n_0$. [For asymmetric oscillations,
described by Eq.~(\ref{TypeI}), we assume that the amplitude is the maximum deviation from the 
equilibrium point.] Note that the parameters $\alpha^{*}$ and $\beta^{*}$ can be either positive or
negative. General periodic solutions for both types of anharmonic oscillators can be given in terms 
of the Jacobi elliptic functions. The corresponding solutions are presented in \ref{AnharmonicOscillator}. 
By substituting these exact solutions into Eq.~(\ref{epsilon-kin}) and making use of the result in 
Eq.~(\ref{E_kin}), we derive 
\begin{equation}
\langle \dot{\cal E}_{\rm diss}\rangle  
=-\frac{\zeta \omega_0^2}{2}\left(\frac {\delta n_0}{n_0}\right)^{2} {\cal F},
\label{zeta-anharmonic}
\end{equation}
where the constant ${\cal F}$ for each type of solution  is determined in terms of  $\alpha^{*}$ and $\beta^{*}$,
see Eqs.~(\ref{FtypeI}) and (\ref{FtypeII}),  respectively. As is easy to check, ${\cal F}\to 1$ in the harmonic 
limit $\alpha^{*}\to 0$ (Type I) or $\beta^{*}\to 0$ (Type II).

By comparing Eqs.~(\ref{diss-energy}) and (\ref{zeta-anharmonic}), we obtain the following 
expression for the bulk viscosity:
\begin{equation}
\zeta=\frac{2 n_0}{\omega_0^2 (\delta n_0)^2 {\cal F}}
\frac{1}{\tau}  \int_0^{\tau} P \delta\dot{n} \, dt .
\label{zeta-pressure}
\end{equation} 
When there is a departure from $\beta$ equilibrium, the pressure can be given in 
terms of the instantaneous composition,
\begin{equation}
P = \bar P + \frac{\partial P}{\partial n} \delta n + n (C_1-C_2)\delta X_e+ nC_1 \delta X_s ,
\label{pressure}
\end{equation}
where $X_e\equiv n_e/n$ and $X_s\equiv n_s/n$ is the electron and strangeness fractions, and $\bar P$ 
is the pressure in equilibrium. The susceptibility functions $C_1$ and $C_2$ were defined in 
Ref.~\cite{Sa'd:2007ud}. By taking into account that $\delta n$ is a periodic function, one finds that only 
the last two terms in the pressure (\ref{pressure}) contribute to the bulk viscosity (\ref{zeta-pressure}),
\begin{equation}
\zeta = \frac{2 n_0^2}{\omega_0^2 (\delta n_0)^2  {\cal F}}
\frac{1}{\tau}  \int_0^{\tau}  \left[ (C_1-C_2)\delta X_e+C_1 \delta X_s\right]  \delta\dot{n} \, dt.
\label{bulk-integral}
\end{equation} 
The instantaneous composition of quark matter is determined by the weak processes, shown 
in Fig.~\ref{fig-weak}. Taking all of them into account, we derive the following set of nonlinear
differential equations for the electron and strangeness fractions:  
\begin{eqnarray}
n\frac{d (\delta X_e)}{dt}&=&\left(\Gamma _d-\Gamma _c\right)+\left(\Gamma _f-\Gamma _e\right) 
=\lambda _2 \delta \mu _2 \sum _{j=0}^2 \chi _j\left(\frac{\delta \mu _2}{T}\right)^{2j} \nonumber \\
&&+\lambda _3\left(\delta \mu _2-\delta \mu _1\right)\sum _{j=0}^2 \chi _j\left(\frac{\delta \mu _2-\delta \mu _1}{T}\right)^{2j} ,
\label{df1}\\
n \frac{d (\delta X_s)}{dt}&=&\left(\Gamma _a-\Gamma _b\right)+\left(\Gamma _c-\Gamma _d\right)\nonumber \\
&=&-\lambda _1 \delta \mu _1 \sum _{j=0}^1 \Upsilon _j\left(\frac{\delta \mu _1}{T}\right)^{2j}
-\lambda _2 \delta \mu _2 \sum _{j=0}^2 \chi _j\left(\frac{\delta \mu _2}{T}\right)^{2j}
\label{df2}
\end{eqnarray}
where $\delta\mu_{1} \equiv \mu_s-\mu_d$, $\delta\mu_{2} \equiv \mu_s-\mu_u-\mu_e$, and the 
notation for the $\lambda$-rates are the same as in Refs.~\cite{Sa'd:2007ud,Wang:2010ydb},
\begin{eqnarray}
\lambda_1 &=&  \frac{64}{5\pi^3}G_F^2 \sin^2\theta_c \cos^2 \theta_c \mu _d^5 T^2, \\
\lambda_2 &=&  \frac{17}{40\pi}G_F^2  \sin^2 \theta_c \mu _s m_s^2 T^4,\\
\lambda_3 &=& \frac{17}{15\pi^2}G_F^2 \cos^2 \theta_c  \alpha_s \mu _d\mu _e\mu_u T^4 .
\end{eqnarray}
In Eqs.~(\ref{df1}) and (\ref{df2}), all higher order corrections in powers of $\delta \mu_i/T$ were taken 
into account, while higher order corrections in powers of $\delta\mu_i/\mu$ were neglected. This is the 
same approximation that was used in Ref.~\cite{Alford:2010gw}. By calculating the semileptonic 
rates using the approach of Ref.~\cite{Sa'd:2006qv}, it is easy to check that the coefficients of the nonlinear 
terms are the same as in the nucleon direct Urca process \cite{Haensel1992,Reisenegger:1994be}: 
$\chi_0 = 1$, $\chi_1 = \frac{10}{17\pi^2}$, and $\chi_2 = \frac{1}{17\pi^4}$. The corresponding 
coefficients in the nonleptonic rates are $\Upsilon_0 = 1$ and $\Upsilon _1 = \frac{1}{4\pi^2}$ 
\cite{Madsen:1992sx,Alford:2010gw,Madsen:1993xx}.

The functions that describe the deviation from equilibrium, $\delta\mu_i$, can be equivalently rewritten in terms 
of the electron and strangeness fractions: $\delta\mu_i = C_i\frac{\delta  n}{n}+ B_i \delta X_e + A_i \delta X_s$,
where coefficient functions $ A_i$, $B_i $ and $C_i$ were defined in Ref.~\cite{Sa'd:2007ud}. By making use of
these relations and Eqs.~(\ref{df1}) and (\ref{df2}), we derive the following self-consistent set of equations for the 
dimensionless quantities $\nu_i\equiv \delta\mu_i /T$ (for $i=1,2$),
\begin{eqnarray}
n_0\frac{d \nu_i}{dt} &=& \frac{C_i}{T}  \delta \dot{n} 
- \lambda _1  A_i \sum _{j=0}^1 \Upsilon _j\left(\nu_1\right)^{2j+1}
- \lambda _2 (A_i-B_i)  \sum _{j=0}^2 \chi _j\left(\nu_2\right)^{2j+1}\nonumber\\
&&+\lambda _3 B_i \sum _{j=0}^2 \chi _j\left(\nu_2-\nu_1\right)^{2j+1} . 
\label{difNu}
\end{eqnarray}
In this study, $\delta n $ is a periodic function that describes anharmonic oscillations of either 
Type~I or Type~II, see Eqs.~(\ref{TypeI}) and (\ref{TypeII}). We make use of the analytical results 
in \ref{AnharmonicOscillator} and solve Eq.~(\ref{difNu}) numerically. When the solutions 
for $\nu_i$ (with $i=1,2$) are available, one can invert the relations for $\delta\mu_i$ in terms 
of $\delta  n$, $\delta X_e$ and $\delta X_s$ in order to determine the deviation of the electron 
and strangeness fractions, 
\begin{eqnarray}
\delta X_e &=& G_e \frac{\delta n}{n_0}+ H_e \nu_1 + J_e \nu_2 ,
\label{deltaX_e} \\
\delta X_s &=& G_s \frac{\delta n}{n_0}+ H_s \nu_1 + J_s \nu_2 .
\label{deltaX_s}
\end{eqnarray}
\label{delta_X}
where 
\begin{eqnarray}
& G_e = \frac{A_1 C_2-A_2 C_1}{A_2 B_1-A_1 B_2}, \qquad \qquad
& G_s = \frac{B_2 C_1-B_1 C_2}{A_2 B_1-A_1 B_2}, \\
& H_e = \frac{T A_2}{A_2 B_1-A_1 B_2},  \qquad \qquad
& H_s = -\frac{T B_2}{A_2 B_1-A_1 B_2}, \\
& J_e = -\frac{T A_1}{A_2 B_1-A_1 B_2},  \qquad \qquad
& J_s = \frac{T B_1}{A_2 B_1-A_1 B_2}.
\end{eqnarray}
Finally, by making use of these results in Eq.~(\ref{bulk-integral}), we can calculate the bulk viscosity,
\begin{eqnarray}
\zeta &=& \frac{2 T n_0^2}{\omega_0^2 (\delta n_0)^2  {\cal F}}
\frac{A_2(C_1-C_2)-B_2 C_1}{A_2 B_1-A_1 B_2}  
\frac{1}{\tau}  \int_0^{\tau} \nu_1 \, \delta\dot{n} \, dt\nonumber\\
&+& \frac{2 T n_0^2}{\omega_0^2 (\delta n_0)^2  {\cal F}}
\frac{B_1 C_1-A_1(C_1-C_2)}{A_2 B_1-A_1 B_2}  
\frac{1}{\tau}  \int_0^{\tau} \nu_2 \, \delta\dot{n} \, dt.
\label{zeta-general}
\end{eqnarray}

\section{Harmonic oscillations}
\label{harmonic-regime}

\begin{figure}[t]
\noindent
\hbox{\includegraphics[width=0.47\textwidth]{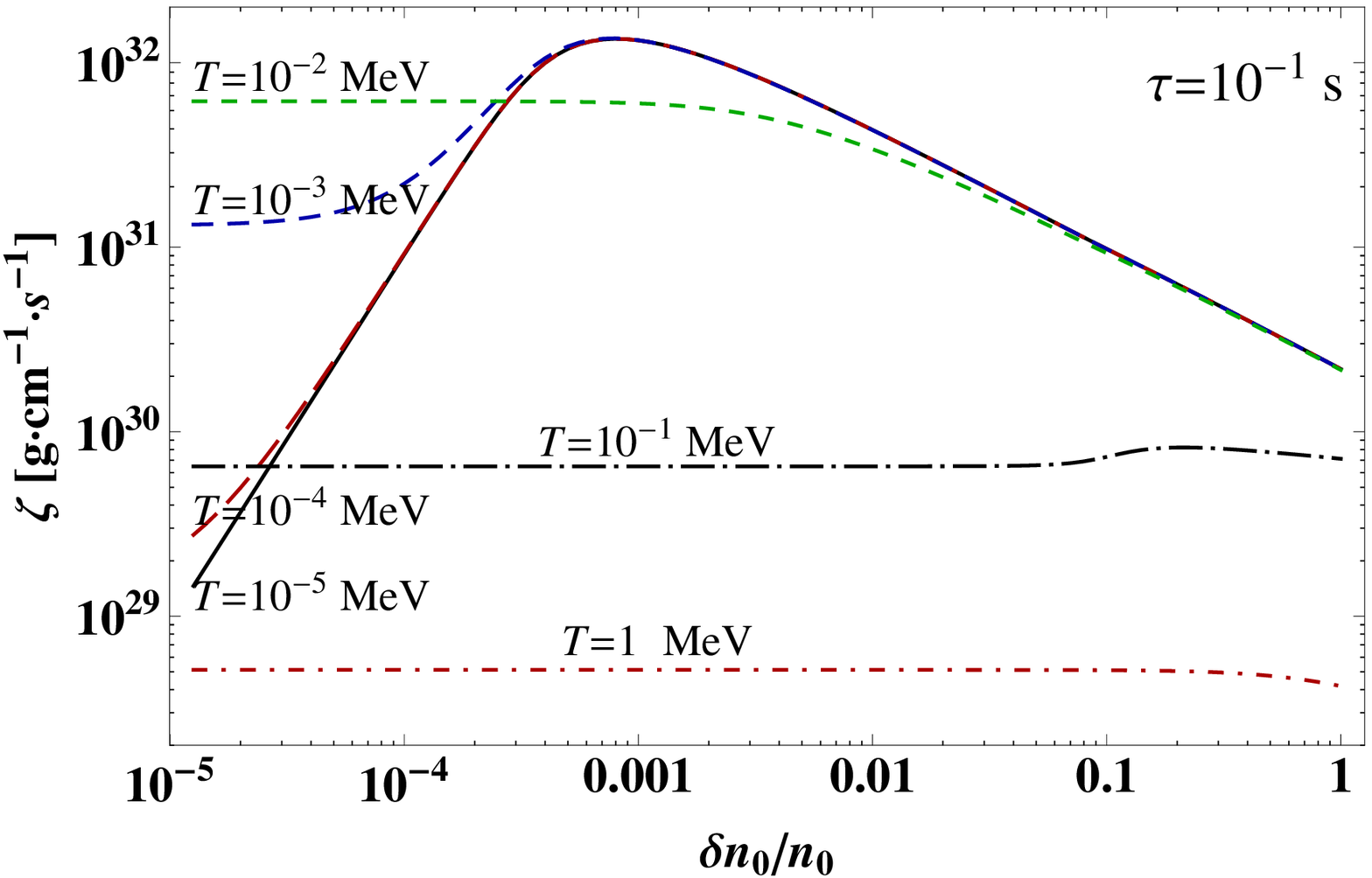}\quad
\includegraphics[width=0.47\textwidth]{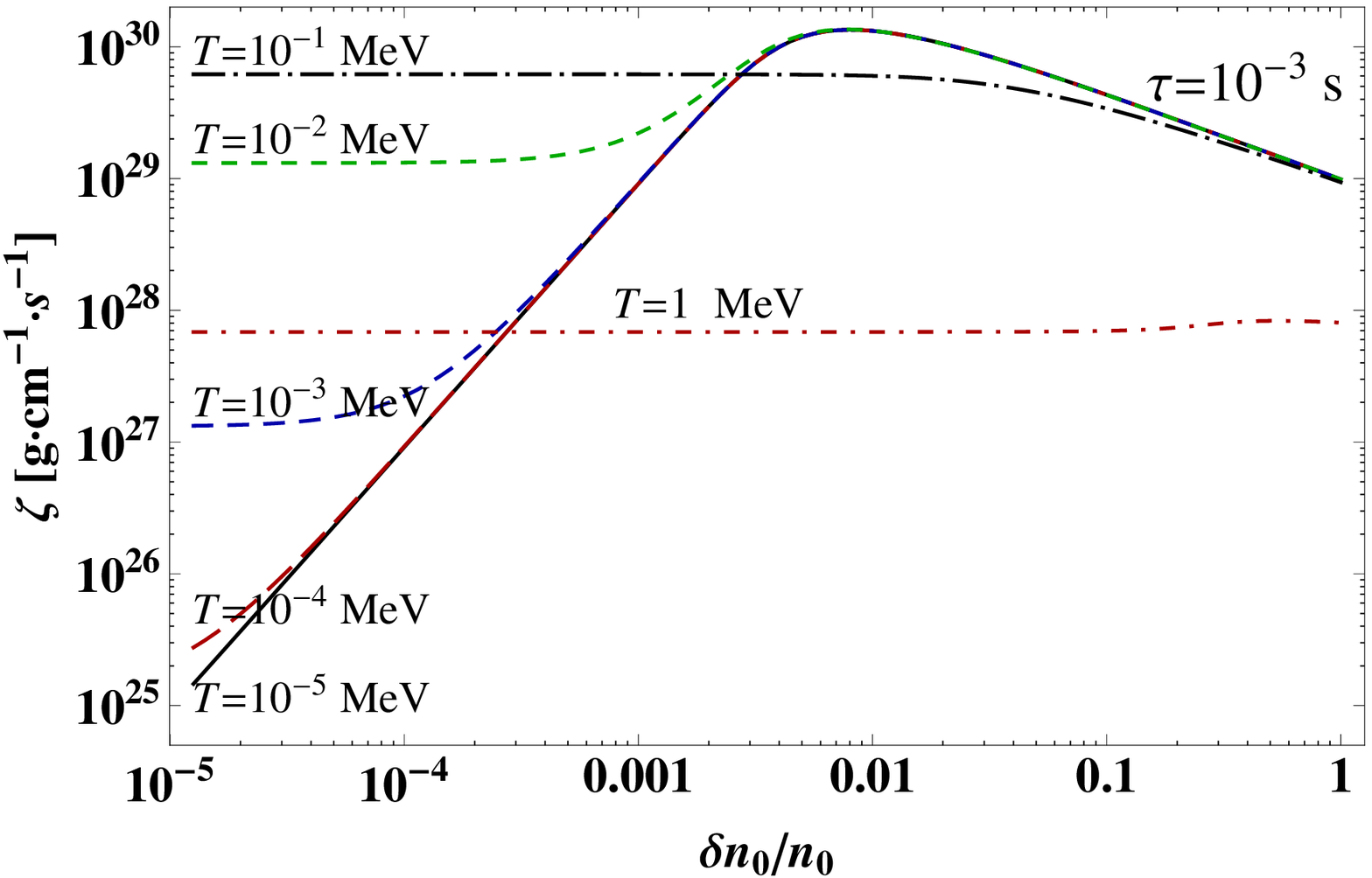}}
\caption{Bulk viscosity as a function of $\delta n_0/n_0$ for several fixed values of  temperature, i.e. 
$T=10^{-5}~\mbox{MeV}$ (black solid line),
$T=10^{-4}~\mbox{MeV}$ (red long-dashed line),
$T=10^{-3}~\mbox{MeV}$ (blue dashed line),
$T=10^{-2}~\mbox{MeV}$ (green short-dashed line),
$T=10^{-1}~\mbox{MeV}$ (black dash-dotted line)
and
$T=1~\mbox{MeV}$ (red dash-dotted line line).}
\label{fig-bulk_vs_deltaN}
\end{figure}

\begin{figure}[t]
\noindent
\hbox{\includegraphics[width=0.47\textwidth]{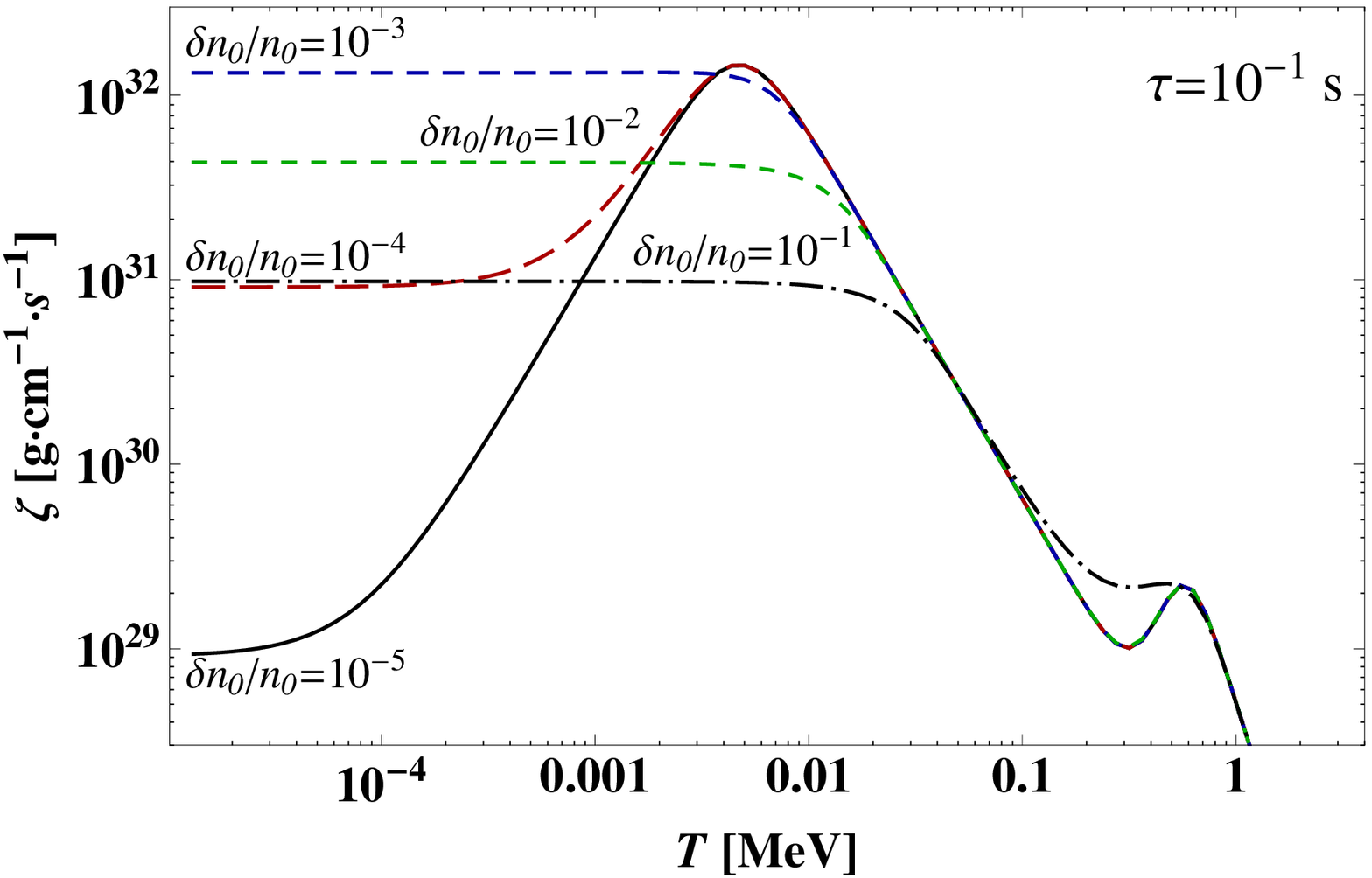}\quad
\includegraphics[width=0.47\textwidth]{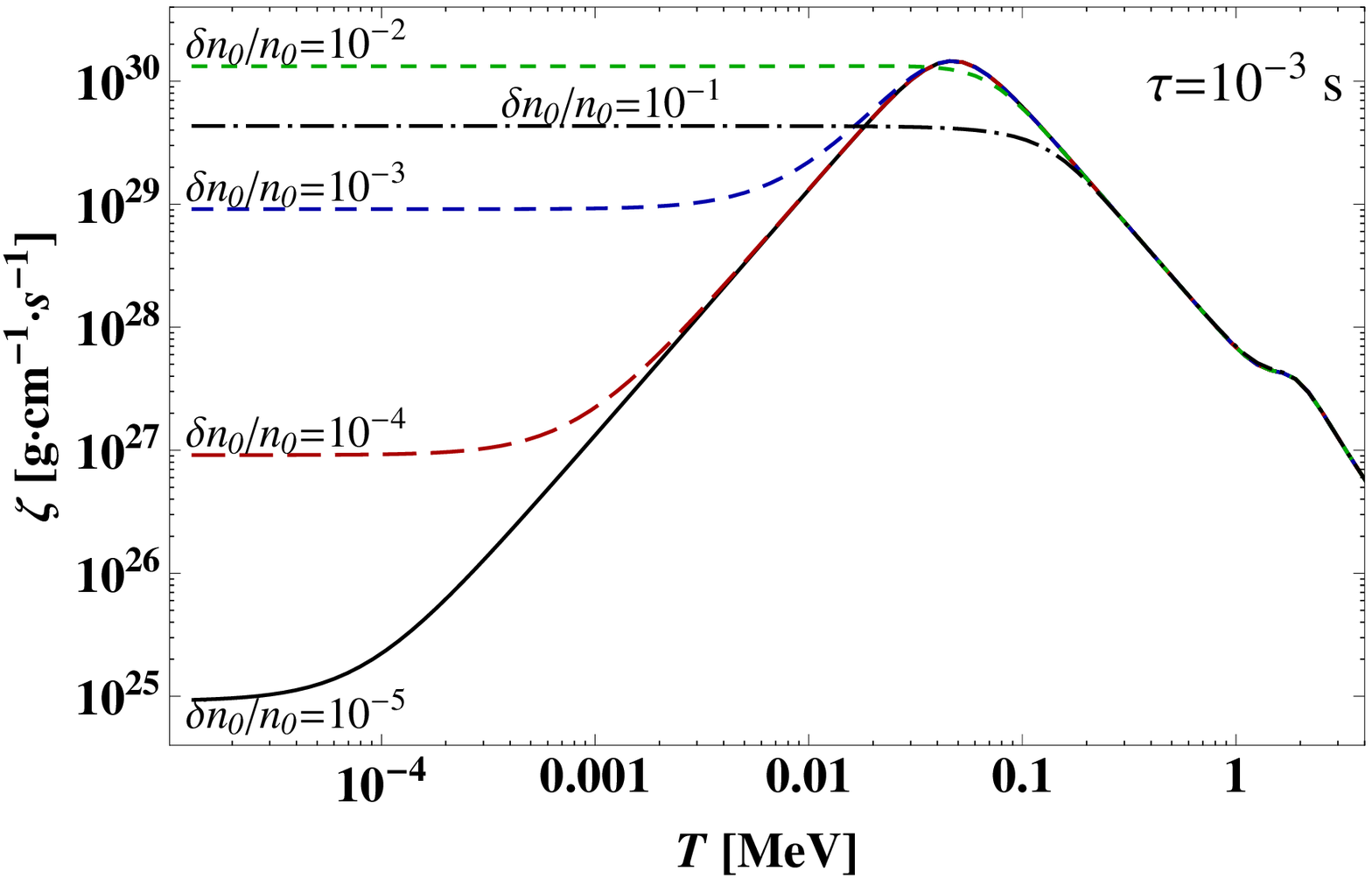}}
\caption{Bulk viscosity as a function of temperature for several fixed values of the amplitude of density oscillations, i.e. 
$\delta n_0/n_0=10^{-5}$ (black solid line),
$\delta n_0/n_0=10^{-4}$ (red long-dashed line),
$\delta n_0/n_0=10^{-3}$ (blue dashed line),
$\delta n_0/n_0=10^{-2}$ (green short-dashed line),
and
$\delta n_0/n_0=10^{-1}$ (black dash-dotted line).}
\label{fig-bulk_vs_T}
\end{figure}

In the limiting case of harmonic oscillations, the density deviations are described by
$\delta n=\delta n_0 \sin(\omega_0 t)$ and the constant ${\cal F}$ in Eq.~(\ref{zeta-anharmonic}) 
is equal to $1$. The numerical results for the bulk viscosity as a function of $\delta n_0/n_0$ 
for several fixed values of temperature are shown in Fig.~\ref{fig-bulk_vs_deltaN}. The linear 
regime corresponds to small values of $\delta n_0/n_0$ , where the bulk viscosity saturates. 
It is also interesting to present the temperature dependence of the bulk viscosity. The corresponding 
plots for several fixed values of $\delta n_0/n_0$ are shown in Fig.~\ref{fig-bulk_vs_T}. As we 
see, with decreasing the temperature, the bulk viscosity eventually levels off. This is the 
outcome of reaching the nonlinear regime, and it would be absent if the linear approximation 
were used instead. In both Fig.~\ref{fig-bulk_vs_deltaN} and Fig.~\ref{fig-bulk_vs_T}, the results for 
two representative values of the period of oscillations, $\tau=0.1~\mbox{s}$ and 
$\tau=10^{-3}~\mbox{s}$, are shown. (In all calculations, we used the model parameters from 
Ref.~\cite{Sa'd:2007ud} at $n=5\rho_0$, where $\rho_0\simeq 0.15~\mbox{fm}^{-3}$ is the 
nuclear saturation density.) In most regions of the parameter space, our results qualitatively 
agree with earlier findings in 
Refs.~\cite{Madsen:1992sx,WangLu,Sawyer,Zheng,Xiaoping,Dong:2007mb,Huang:2009ue,Alford:2010gw}. 
The only notable difference occurs in Fig.~\ref{fig-bulk_vs_T} around the ``semileptonic" hump 
($T\sim 1~\mbox{MeV}$). This comes from the interplay of semileptonic weak processes with the 
more dominant nonleptonic ones \cite{Sa'd:2007ud,Wang:2010ydb}. In most of the studies, this is 
neglected because of a smallness of the semileptonic rates.


Here it is appropriate to mention that, in application to stellar quark matter, the bulk viscosity may not 
be the only, or even the dominant mechanism responsible for damping of the r-mode instabilities. 
For example, at sufficiently low temperatures, the corresponding dissipative dynamics is known to be 
dominated by the shear viscosity \cite{Madsen:1999ci,Jaikumar:2008kh}. (For several representative 
studies of the shear viscosity in dense quark matter see, for example, Refs.~\cite{Heiselberg:1993cr,
Manuel:2004iv,Alford:2009jm}.) In this connection, our low-temperature results in Figs.~\ref{fig-bulk_vs_deltaN}
and \ref{fig-bulk_vs_T} should be used only for the purpose of determining where exactly the damping by 
the shear viscosity takes over. 

In order to better understand the role of nonlinear terms in the weak rates, see Eqs.~(\ref{df1}) and 
(\ref{df2}), as well as their effect on the bulk viscosity, it is instructive to study the linear approximation.
In this case, the expression for the viscosity can be derived analytically. As we shall see, this will 
be also helpful to elucidate the role of induced oscillations of $\delta\mu_{2} = \mu_s-\mu_u-\mu_e$ 
when the semileptonic processes are formally switched off.

\subsection{Harmonic oscillations: linear approximation}

In the linear regime, Eqs.~(\ref{df1}) and (\ref{df2}) for dimensionless functions $\nu_i$ simplify down to
\begin{eqnarray}
\frac{d \nu_1}{d \vartheta} &=& d_1 \cos\vartheta - f_1  \nu_1  - f_2 \nu_2 + f_3(\nu_2-\nu_1),
\label{linear1}\\
\frac{d \nu_2}{d \vartheta} &=& d_2 \cos\vartheta - h_1 \nu_1 - h_2 \nu_2 + h_3 (\nu_2-\nu_1) ,
\label{linear2}
\end{eqnarray}
where $\vartheta = 2\pi t/\tau$ is the dimensionless time variable, $d_i  = C_i \delta n_0/(Tn_0)$ is 
the magnitude of the ``driving force", and the other coefficient functions are
\begin{eqnarray}
& f_1 = \lambda_{1}\frac{A_{1}}{\omega_0 n_0},  \qquad \qquad
& h_1 =  \lambda_{1}\frac{A_{2} }{\omega_0 n_0} ,\\
& f_2 = \lambda_{2}\frac{A_{1}-B_{1}}{\omega_0 n_0},  \qquad \qquad
& h_2 = \lambda_{2} \frac{A_{2} -B_{2}}{\omega_0 n_0} ,\\
& f_3 = \lambda_{3}\frac{B_{1} }{\omega_0 n_0},  \qquad \qquad
& h_3 = \lambda_{3} \frac{B_{2} }{\omega_0 n_0} .
\end{eqnarray}
The general solution to this set of equations in the steady state regime is given by 
\begin{eqnarray}
 \nu_1 &=& x_1 \cos\vartheta +y_1 \sin\vartheta,\\
 \nu_2 &=& x_2 \cos\vartheta +y_2 \sin\vartheta  ,
\end{eqnarray}
where the coefficients satisfy the following set of algebraic equations:
\begin{eqnarray}
(f_1+f_3)x_1  +(f_2-f_3) x_2 +y_1 &=& d_1 ,\\
-x_1 +(f_1+f_3) y_1+(f_2-f_3) y_2 &=& 0 ,\\
(h_1+h_3)x_1  +(h_2-h_3) x_2 +y_2 &=& d_2 ,\\
-x_2 +(h_1+h_3)y_1+(h_2-h_3)y_2&=& 0.
\end{eqnarray}
It is straightforward, although tedious to solve this set of equations. When the solution is available, the 
result for the bulk viscosity will follow from Eq.~(\ref{zeta-general}), i.e.
\begin{eqnarray}
\label{bulk}
\zeta &=& \frac{T n_0^2}{\omega \delta n_0} \frac{A_2(C_1-C_2)-B_2 C_1}{A_2 B_1-A_1 B_2}  x_1
               +\frac{T n_0^2}{\omega \delta n_0} \frac{B_1C_1-A_1(C_1-C_2)}{A_2 B_1-A_1 B_2}  x_2 .
\end{eqnarray} 
It can be shown that this expression (with the appropriate solutions for $x_1$ and $x_2$) 
coincides exactly with the result for the bulk viscosity, obtained in Ref.~\cite{Sa'd:2007ud}.

\subsection{Harmonic oscillations: nonleptonic contribution in linear approximation}

If one ignores the semileptonic processes (i.e. if one formally takes $\lambda_2=\lambda_3=0$), 
the linearized equations (\ref{linear1}) and (\ref{linear2}) for $\nu_i$'s take the following form:
\begin{eqnarray}
\frac{d \nu_1}{d \vartheta} &=& d_1 \cos\vartheta - f_1  \nu_1  ,\\
\frac{d \nu_2}{d \vartheta} &=& d_2 \cos\vartheta - h_1 \nu_1 .
\end{eqnarray}
The explicit solution to this set of equations in the steady state regime is given by 
\begin{eqnarray}
 \nu_1 &=& \frac{d_1 }{1+f_1^2}\left( f_1 \cos\vartheta + \sin\vartheta \right),\\
 \nu_2 &=& d_2 \sin\vartheta +\frac{d_1 h_1}{1+f_1^2} \left(\cos\vartheta - f_1 \sin\vartheta \right)  ,
\end{eqnarray}
and the corresponding expression for the bulk viscosity reads
\begin{eqnarray}
\zeta_{\rm non} &=& \frac{T n_0^2}{\omega \delta n_0} \frac{A_2(C_1-C_2)-B_2 C_1}{A_2 B_1-A_1 B_2}  
\frac{d_1 f_1 }{1+f_1^2} \nonumber\\
&+&\frac{T n_0^2}{\omega \delta n_0} \frac{B_1 C_1-A_1(C_1-C_2)}{A_2 B_1-A_1 B_2}  
\frac{d_1 h_1}{1+f_1^2}=\frac{\lambda_1 C_1^2}{\omega^2+(\lambda_1 A_1/n_0)^2},
\end{eqnarray} 
where we used the relation $f_1/h_1=A_1/A_2$ to arrive at the final result. As expected, this 
agrees with the known result \cite{Madsen:1992sx,Sawyer,Alford:2010gw,Sa'd:2007ud}.

It is interesting to notice that the final result for the bulk viscosity receives a nonzero contribution due to the 
oscillation of $\nu_2 \equiv \delta \mu_2/T$. Since $\delta \mu_2\equiv  \mu_s-\mu_u-\mu_e$ controls the 
imbalance of the rates in the semileptonic processes, shown in Figs.~\ref{fig-weak} $(c)$ and $(d)$, which are 
formally switched off in the approximation at hand, one might wonder why there should be such a contribution
at all. The answer is quite simple. In absence of the semileptonic processes the electron fraction in quark matter 
cannot change. However, when the nonleptonic processes drive the oscillations of the strangeness composition, 
they inevitably induce the oscillations of $\delta \mu_2\equiv  \mu_s-\mu_u-\mu_e$. Then, the latter contributes 
to the instantaneous pressure and, in turn, to the bulk viscosity. Interestingly, such a contribution 
due to the induced oscillation of $\nu_2 \equiv \delta \mu_2/T$ were ignored in all previous studies 
\cite{Madsen:1992sx,WangLu,Sawyer,Zheng,Xiaoping,Dong:2007mb,Huang:2009ue,Alford:2010gw}. Fortunately, 
the corresponding correction is quantitatively small. The reason for its smallness seems to be rooted in the
 ``accidental" fact that one of the susceptibility functions, $B_2$, is inversely proportional to the square of 
 the chemical potential of electrons (rather than quarks) and, thus, is considerably larger than all others 
 \cite{Sa'd:2007ud}.

\section{Bulk viscosity in anharmonic regime}
\label{anharmonic-regime}

In this section we study the effect that anharmonic oscillations of the density have on the bulk viscosity 
of dense quark matter. 

\subsection{Anharmonic oscillations of Type I}

Let us start by modeling the density oscillations of quark matter $\delta n(t)$ by a time dependent anharmonic 
function of Type I, which is a solution to Eq.~(\ref{TypeI}) with a fixed anharmonicity parameter $\alpha^{*}$. 
Before we proceed to the numerical results, it is important to notice that the corresponding oscillations are 
asymmetric with respect to the equilibrium point $\delta n_{\rm eq}=0$. For $\alpha^{*}<0$, the density 
oscillations are larger in the direction of positive $\delta n$, while for $\alpha^{*}>0$, the oscillations are 
larger in the direction of negative $\delta n$, see also Fig.~\ref{fig-U3}. The cases of the positive and negative
anharmonicity parameters are physically equivalent, however. Indeed, they are related by the following sign 
reversal symmetry: $\alpha\to -\alpha$ and $\delta n\to -\delta n$. Therefore, it is sufficient to study only 
one of them. For technical reasons, we choose $\alpha^{*}<0$. 

\begin{figure}[t]
\noindent
\hbox{\includegraphics[width=0.46\textwidth, bb = 12 12 320 232]{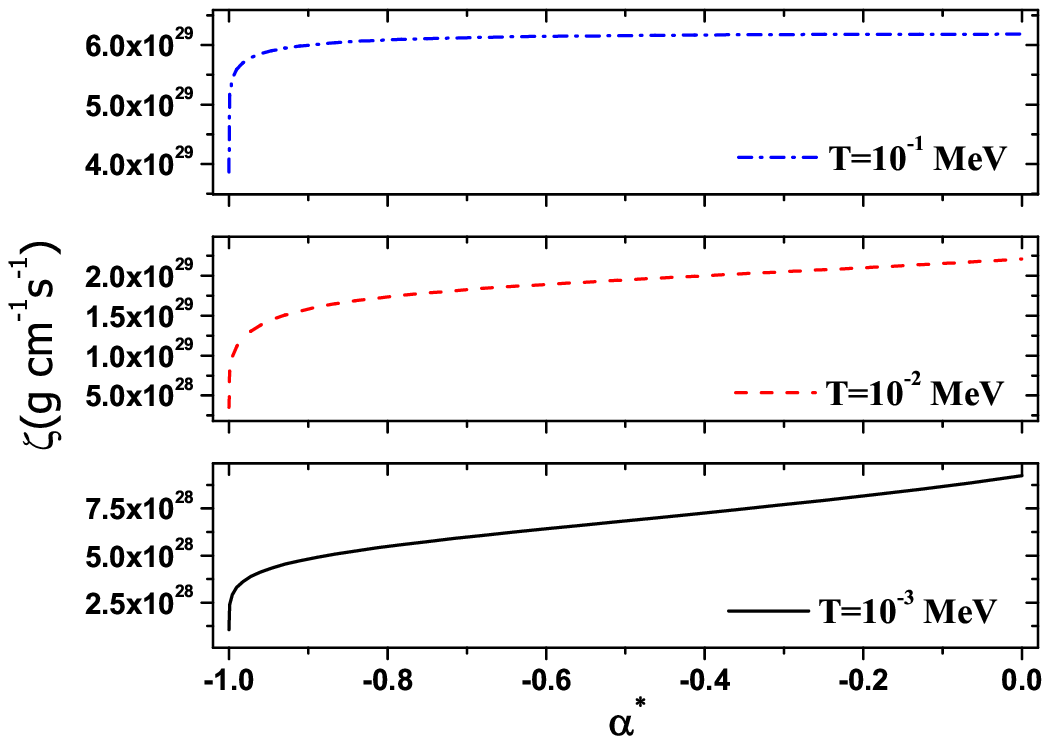}\,
\includegraphics[width=0.46\textwidth, bb = 12 12 320 232]{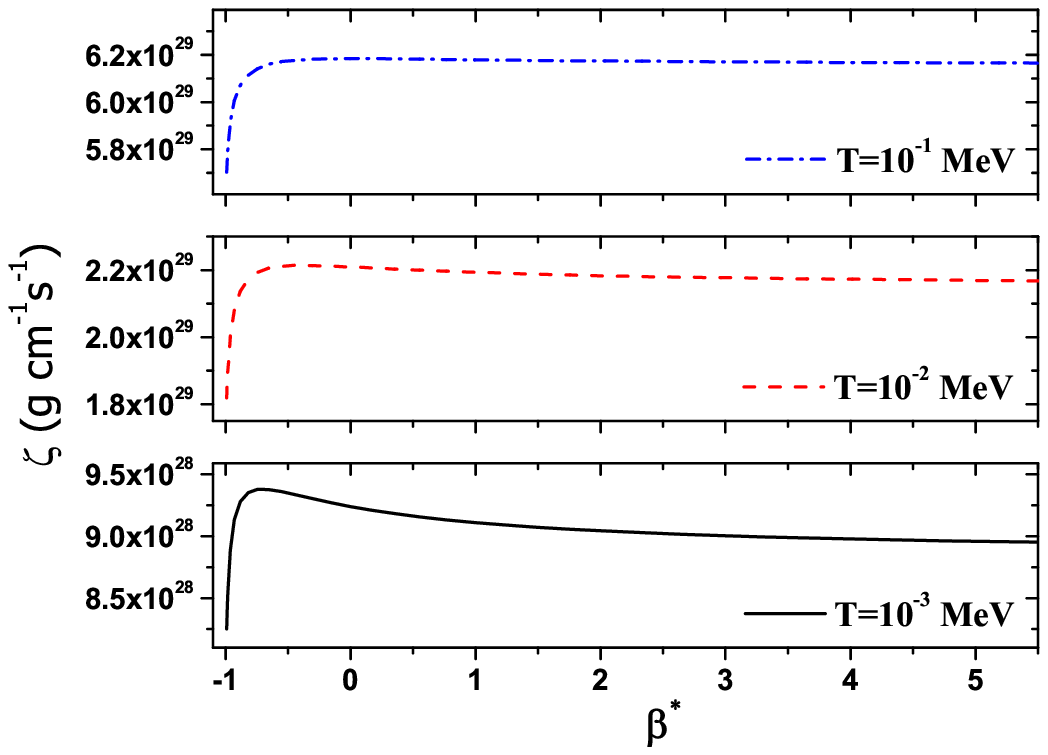}}
\caption{Bulk viscosity as a function of anharmonicity parameter of Type I (left panel) and 
Type II (right panel) for $\tau =10^{-3}~\mbox{s}$, $\delta n_0/n_0=10^{-3}$ and several 
representative values of temperature.}
\label{fig-bulk_anharmonic}
\end{figure}

Typical results for the bulk viscosity as a function of anharmonicity parameter $\alpha^{*}$ are shown in the
left panel of Fig.~\ref{fig-bulk_anharmonic} for the whole range of negative $\alpha^{*}$, i.e. $-1 <\alpha^{*}<0$, 
for which physically meaningful periodic solutions exist. We used the following values of the period and the amplitude 
of density oscillations: $\tau =10^{-3}~\mbox{s}$ and $\delta n_0/n_0=10^{-3}$, and plotted the results
for several representative values of temperature. [It should be emphasized that the period $\tau$ is related to
the ``bare" frequency $\omega_0$ by a modified relation, see Eq.~(\ref{tau-alpha}).] In general, 
we find that the bulk viscosity decreases with 
increasing the degree of anharmonicity. This qualitative behavior may be understood as the result of an 
effective increase of the frequency of oscillations due to an admixture of higher harmonics. Quantitatively,
however, the effect is rather small. Only a very large anharmonicity ($\alpha^{*}\approx -1$) leads to a substantial 
decrease of the bulk viscosity.

\subsection{Anharmonic oscillations of Type II}

Anharmonic density oscillations of Type II are modeled by a function $\delta n(t)$, which is a solution to 
Eq.~(\ref{TypeII}) with a fixed anharmonicity parameter $\beta^{*}$. Conceptually, this is a simpler case
because the oscillations are symmetric about the equilibrium point $\delta n_{\rm eq}=0$. Unlike the case 
of Type I oscillations, there is no reversal symmetry here. As in the previous case, however, periodic solution
exist only for a range of values of the the anharmonicity parameter, $\beta^{*}>-1$. 

Numerical results for the bulk viscosity as a function of anharmonicity parameter $\beta ^{*}$ are shown in the
right panel of Fig.~\ref{fig-bulk_anharmonic}. The qualitative dependence of the viscosity on the parameter 
 $\beta ^{*}$ is somewhat different. While it decreases at large values of parameter $\beta ^{*}$, there is a
range of small negative values of $\beta ^{*}$, where it slightly grows with increasing anharmonicity.  Moreover,
this feature seems to be rather general and especially pronounced in the nonlinear regime (small temperature).
Just like in the case of Type I oscillations, the effects appear to be rather small. 

\section{Discussion}
\label{discussion}

In this paper we studied the bulk viscosity of dense quark matter by taking into account the nonlinear 
dependence of the nonleptonic and semileptonic weak rates on the parameter $\delta\mu_i /T$,
where $\delta\mu_i $ are the chemical potential imbalances that control the departure of strange quark matter
from $\beta$ equilibrium. We reproduce the earlier obtained interplay of the nonleptonic and 
semileptonic processes, leading to an increase (``hump") of the viscosity in a narrow temperature 
range around $1~\mbox{MeV}$. The nonlinear corrections have a small effect on the 
corresponding shape of the ``hump". The reason for this is a relatively high temperature ($T\sim 
1~\mbox{MeV}$), at which the corresponding effects can occur. At such moderately high 
temperatures, the interplay between the two types of weak processes is substantially affected 
only if the nonlinearity (measured by $\delta n_0/n_0$) is well above $10\%$. 

In this study, we also found that the anharmonicity of density oscillations has an effect on 
the bulk viscosity, even though the effect was not large in the cases that we studied. For
a strong anharmonicity, the bulk viscosity showed a substantial decrease.  We also saw 
that different types of anharmonicity have slightly different qualitative as well as quantitative 
outcomes. This finding may suggest that some types of anharmonicity may be more efficient 
and, thus, lead to larger corrections to the bulk viscosity. In this preliminary study, we 
considered only a toy model to introduce the anharmonic oscillations. In the future, it may 
be interesting to study more realistic types of density oscillations that result from the actual 
nonlinear dynamics of stellar r-modes, produced by the gravitational emission.

Here we studied the bulk viscosity only in the normal phase of dense quark matter. 
One may wonder, however, how the results will be modified if quark matter is a color 
superconductor \cite{Alford:2007xm}. As suggested by several existing studies in the 
linear regime, color superconductivity can have a large effect on the bulk viscosity 
\cite{Wang:2010ydb,Sa'd:2006qv,Alford:2006gy,Alford:2008pb}. The additional effects due to 
the nonlinear regime may be much harder to predict. One of the complications comes from the 
fact that the weak rates in color superconducting matter have a strong dependence on the 
ratio of the superconducting energy gap and temperature, $\phi/T$. Moreover, in the suprathermal 
regime, their dependence on $\delta\mu_i/T$ will most likely be nonpolynomial. These 
technical difficulties can be resolved in principle, but the outcome is not obvious and the 
corresponding study is outside the scope of the present paper.

\ack
\addcontentsline{toc}{section}{Acknowledgments}
This work is supported in part by the start-up funds from the Arizona State University and by the 
U.S. National Science Foundation under Grant No. PHY-0969844.

\appendix

\section{Anharmonic oscillator}
\label{AnharmonicOscillator}

\subsection{Anharmonic oscillator with a cubic potential (Type I)}

\begin{figure}[t]
\noindent
\hbox{\includegraphics[width=0.48\textwidth]{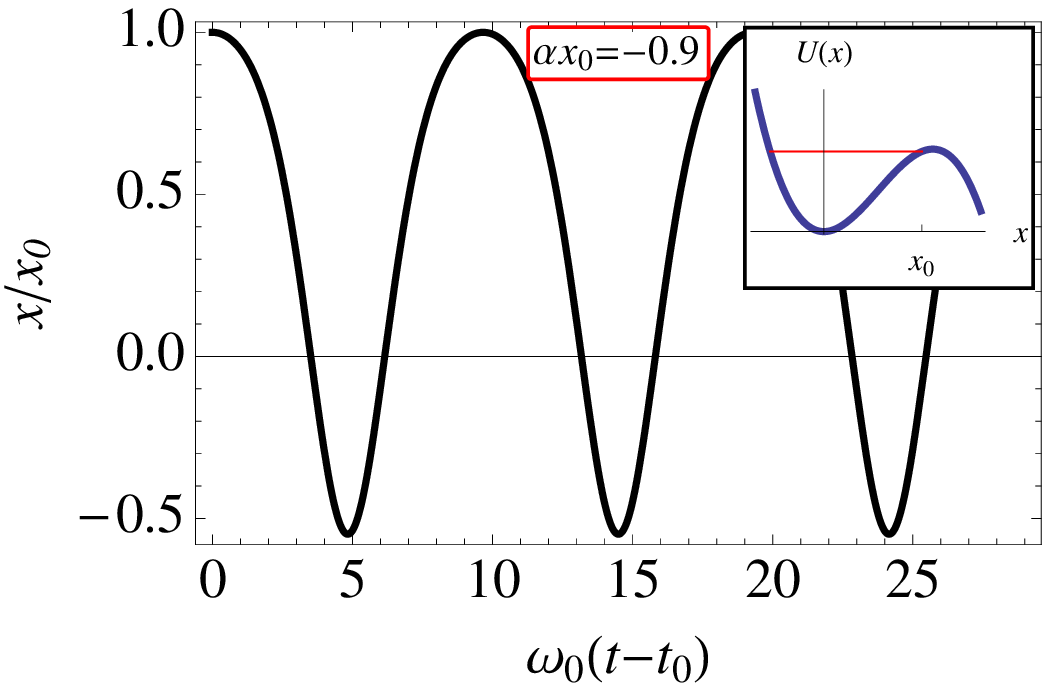}\quad
\includegraphics[width=0.48\textwidth]{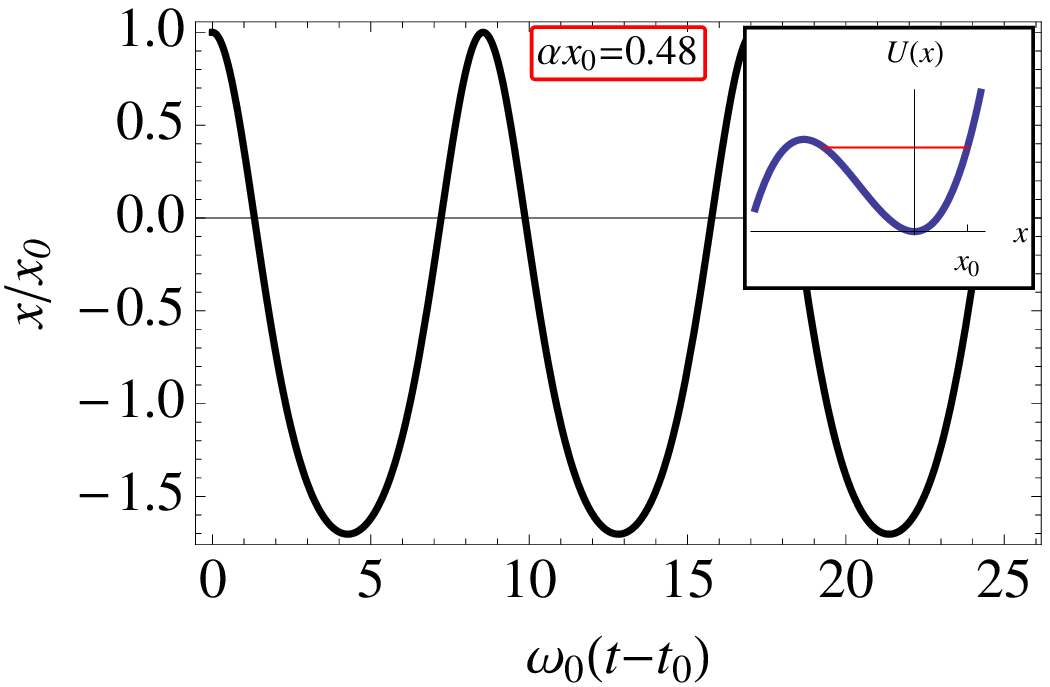}}
\caption{Solutions to the equation of motion of the anharmonic oscillator with a cubic potential for two values of 
the coupling constant. The inserts show the corresponding shapes of the potentials.}
\label{fig-U3}
\end{figure}

The anharmonic oscillator with cubic potential 
$U(x)=m \omega_0^2 \left(\frac{x^2}{2}+\alpha\frac{x^3}{3}\right) $ 
is described by the following equation of motion:
\begin{equation}
\ddot{x} +\omega_0^2 x \left(1+\alpha x\right) =0.
\end{equation}
By making use of the known parametric solution to the above differential equation \cite{DiffEqs} 
and assuming that $x_0>0$ is the maximum deviation from the equilibrium point $x=0$ 
in the positive $x$-direction, we find that the periodic solution exists for $-1<\alpha x_0<1/2$. 
(The apparent asymmetry between positive and negative values of $\alpha x_0$ is a result of the
assumption that $x_0$ is the maximum deviation from the equilibrium point in the {\em positive} 
$x$-direction.) It is given in terms of the Jacobi elliptic function $\mbox{sn} \left(u|m\right)$ as follows:
\begin{equation}
x(t) = x_0
\left[1-\frac{3a_{-} }{2\alpha^{*}} \mbox{sn} \left( \frac{(t-t_0)\omega_0 \sqrt{a_{+}}}{2}\Big|\frac{a_{-}}{a_{+}}\right)^2\right],
\end{equation}
where $a_{\pm}=\frac{1}{2}\left(1+2\alpha^{*} \pm \sqrt{(1+\frac{2}{3}\alpha^{*})(1-2\alpha^{*})}\right)$ 
and $\alpha^{*}\equiv\alpha x_0$ is a dimensionless parameter that measures the maximum
deviation of the solution from the harmonic regime.  Note that $x(t)$ is periodic with the period given by
\begin{eqnarray}
\tau_\alpha &=& \frac{4}{\omega_0\sqrt{a_{+}}} K\left(\frac{a_{-}}{a_{+}}\right)\nonumber\\
&\simeq & \frac{2\pi}{\omega_0}\left(1+\frac{5}{12}(\alpha\, x_0)^2
+\frac{5}{18}(\alpha\, x_0)^3+O\left[(\alpha\, x_0)^4\right]\right),~ \mbox{for} ~~ \alpha\, x_0\to 0,
\label{tau-alpha}
\end{eqnarray}
where $K\left(a_{-}/a_{+}\right)$ is the complete elliptic integral of the first kind. Two representative 
solutions are shown in Fig.~\ref{fig-U3}. 
For anharmonic solutions of this type, the average kinetic energy is given by
\begin{equation}
\langle E_{\rm kin} \rangle =\frac{1}{\tau_\alpha}\int_{0}^{\tau_\alpha} \frac{m\dot{x}^2}{2} dt 
=\frac{m\omega_0^2 x_0^2}{4} {\cal F} ,
\label{E_kin}
\end{equation}
where the analytical expression for constant ${\cal F} $ reads
\begin{equation}
{\cal F}  = \frac{4(1+\alpha^{*})}{5\alpha^{*} a_{-}}\left\{
\frac{E\left(\frac{a_{-}}{a_{+}}\right)}{K\left(\frac{a_{-}}{a_{+}}\right)} - (a_{+}-a_{-})\left(a_{+}-\frac{1}{2}a_{-}\right)
\right\}. 
\label{FtypeI}
\end{equation}
Here $E\left(a_{-}/a_{+}\right)$ is the complete elliptic integral of the second kind.

\subsection{Anharmonic oscillator with a quartic potential (Type II)}

\begin{figure}[t]
\noindent
\hbox{\includegraphics[width=0.48\textwidth]{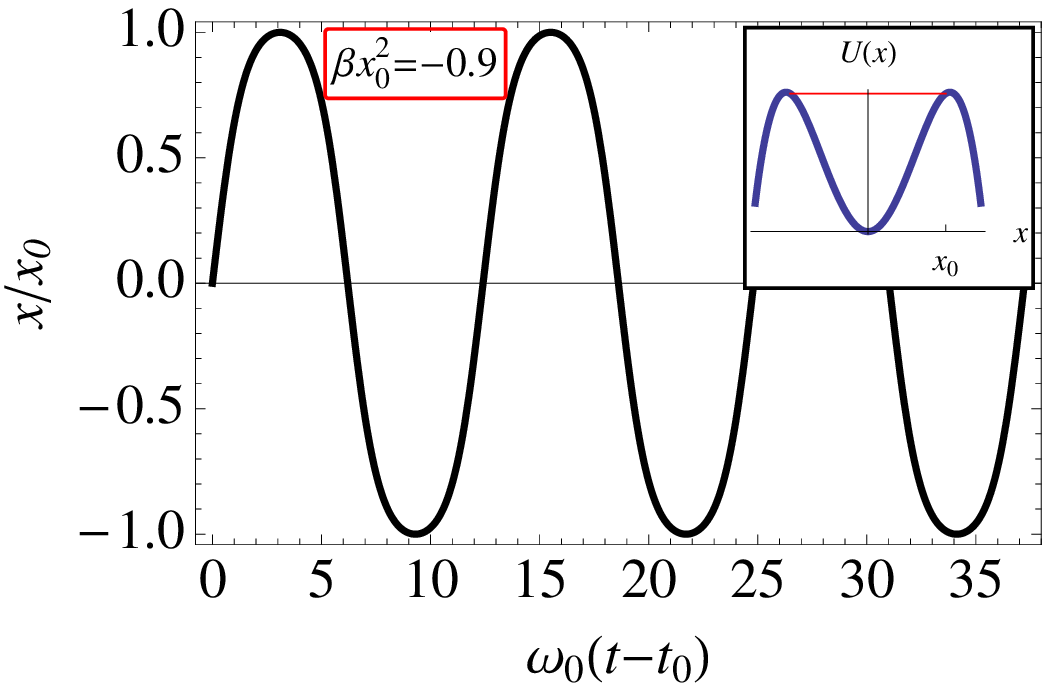}\quad
\includegraphics[width=0.48\textwidth]{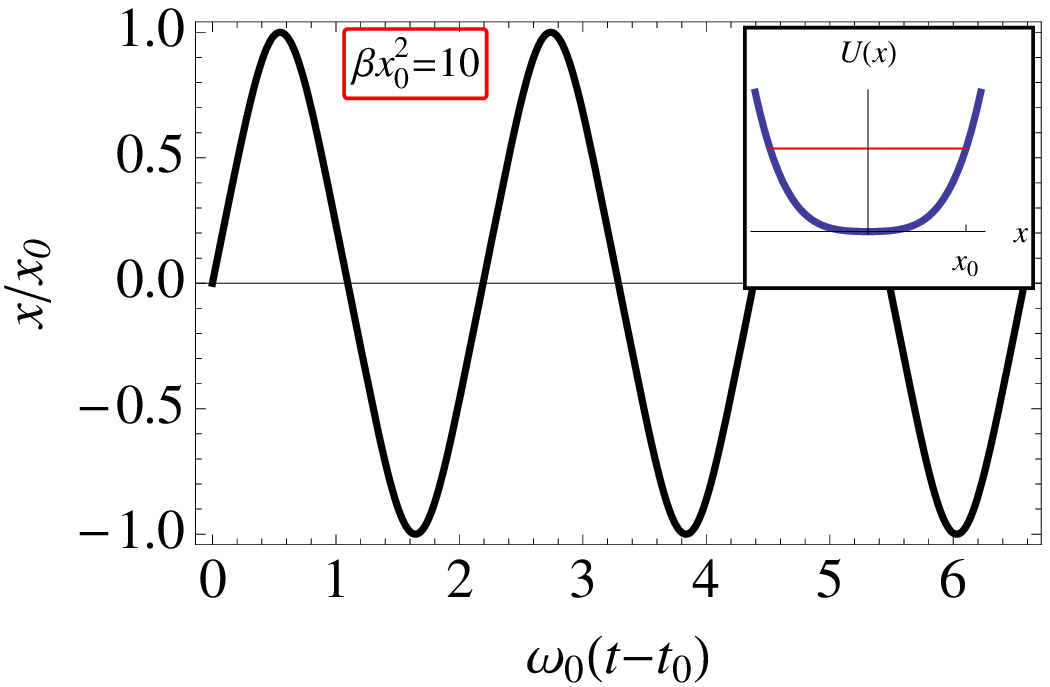}}
\caption{Solutions to the equation of motion of the anharmonic oscillator with a quartic potential for two values of 
the coupling constant. The inserts show the corresponding shapes of the potentials.}
\label{fig-U4}
\end{figure}

The anharmonic oscillator with quartic potential 
$U(x)=m \omega_0^2 \left(\frac{x^2}{2}+ \beta \frac{x^4}{4}\right) $ 
is described by the following equation of motion:
\begin{equation}
\ddot{x} +\omega_0^2 x \left(1+\beta x^2 \right) =0  , \qquad  (\mbox{Type II}).
\end{equation}
Using the known parametric solution to the above differential equation \cite{DiffEqs}
and assuming that $x_0$ is the maximum deviation from the equilibrium point $x=0$, we 
find that the periodic solution exists for $\beta^{*}\equiv \beta\, x_0^2 >-1$. 
The solution is given in terms of the Jacobi elliptic function,
\begin{eqnarray}
x(t)&=& x_0 \frac{\sqrt{1+ (1/2)\beta^{*}}}{\sqrt{1+\beta^{*}}} \, \mbox{sd} \left(\omega_0 (t-t_0) 
\sqrt{1+\beta^{*}} \Big|\frac{\beta^{*}}{2(1+\beta^{*})} \right).
\end{eqnarray}
This is a periodic solution with the period equal
\begin{eqnarray}
\tau_\beta &=& \frac{4}{\omega_0 \sqrt{1+\beta^{*}}}\, K\left(\frac{\beta^{*}}{2(1+\beta^{*})}\right)\nonumber\\
&\simeq & \frac{2\pi}{\omega_0}\left(1-\frac{3}{8}\beta\, x_0^2+\frac{57}{256}\beta^2\, x_0^4+O\left(\beta^3\, x_0^6\right)
\right),\quad \mbox{for} \quad \beta\, x_0^2\to 0.
\end{eqnarray}
Two representative solutions are shown in Fig.~\ref{fig-U4}. 
For solutions of this type, the average kinetic can be given in the same form as in Eq.~(\ref{E_kin}), 
but the value of the corresponding constant ${\cal F} $ is different, 
\begin{equation}
{\cal F}  =  \frac{4(1+\beta^{*})}{3\beta^{*}}\left\{1+\frac{\beta^{*}}{2}
-\frac{E\left(\frac{\beta^{*}}{2(1+\beta^{*})} \right)}{K\left(\frac{\beta^{*}}{2(1+\beta^{*})}\right)}\right\}.
\label{FtypeII}
\end{equation}

\end{document}